# Field induced phase transitions and anisotropic magnetic properties of the Kiteav-Heisenberg compound $Na_2Co_2TeO_6$

A. K. Bera and S. M. Yusuf *

*Solid State Physics Division, Bhabha Atomic Research Centre, Mumbai 400085, India*
*Homi Bhabha National Institute, Anushaktinagar, Mumbai 400094, India*

F. Orlandi and P. Manuel

*ISIS Facility, STFC Rutherford Appleton Laboratory, Harwell Oxford, Didcot OX11 0QX, United Kingdom*

L. Bhaskaran and S. A. Zvyagin

*Dresden High Magnetic Field Laboratory (HLD-EMFL), Helmholtz-Zentrum Dresden-Rossendorf, 01328 Dresden, Germany*

*e-mail: smyusuf@barc.gov.in*

**Abstract:**

Spin systems with honeycomb structures have recently attracted a great deal of attention in connection with the Kitaev quantum spin liquid state (QSL) predicted theoretically. One possible Kitaev QSL candidate is $Na_2Co_2TeO_6$ realizing a honeycomb lattice of pseudo-spin-1/2. Field-dependent single-crystal neutron diffraction technique allows us to determine the microscopic spin-spin correlations across the field induced phase transitions for $H//a$ and $H//a^*$ in plane field directions. Our results reveal phase transitions, initially to a canted zigzag antiferromagnetic state at approximately 60 kOe, followed by a possible transition to a partially polarized state over the range of 90-120 kOe, and finally to a field-induced fully polarized state above 120 kOe. We observe distinct field dependencies of the magnetic peak intensities for $H//a$ and $H//a^*$. In addition, low-temperature electron spin resonance in magnetic fields $H//c$ yields a complete softening for one of the antiferromagnetic resonances at ~ 40 kOe, revealing a field-induced phase transition. The present work, thus, provides new insights into the field evolution of the important Kitaev-Heisenberg spin system $Na_2Co_2TeO_6$.

*Introduction:* The exactly solvable Kitaev honeycomb model [1], based on an effective spin-1/2 two dimensional honeycomb lattice, offers a topological quantum spin liquid (QSL) state with Majorana fermions as excitations. Searching for new materials, where the Kitaev model is realized, is at forefront of modern material science and condensed matter physics. Kitaev interactions or bond-dependent Ising interactions may arise from a spin-orbit entangled pseudospin-1/2 degrees of freedom in transition metal ions located in edge-shared octahedral crystal fields. A honeycomb-lattice arrangement of such ions is an important prerequisite for the realization of Kitaev model [2], as it is believed to be the case of $Na_2IrO_3$ [3] and $\alpha$-$RuCl_3$ [4] with the $d^5$ spin configurations. However, a pure Kitaev material remains elusive due to the presence of additional non-Kitaev interactions viz., Heisenberg exchange interactions between the pseudo spins. Kitaev physics is also predicted for $d^7$ $Co^{2+}$ ions with a high-spin $t_{2g}^5e_g^2$ configuration that can provide pseudospin-1/2 degrees of freedom [5, 6]. In this regard, several Co-based candidate Kitaev materials for instance, $Na_2Co_2TeO_6$ [7-19], $BaCo_2(XO_4)_2$ ($X$= P and As) [20-22], and $A_3Co_2SbO_6$ (with $A$ = Li, Na, and Ag) [12, 23, 24] have been recently suggested. Among them, the layered honeycomb magnet $Na_2Co_2TeO_6$ is of our present interest. Although at low temperature ($T_N$ ~ 25 K) $Na_2Co_2TeO_6$ orders magnetically [10, 11, 15, 16] (similar to $Na_2IrO_3$ [3] and α-$RuCl_3$ [4]), this material has shown some signatures (viz., field induced disappearance of the peaks observed on specific heat and χ(T), as well as enhancement of magnetic entropy with applied magnetic field) of the Kitaev-Heisenberg physics [12, 19].

The crystal structure of $Na_2Co_2TeO_6$ [Fig. 1(a)] (hexagonal symmetry, space group $P6_322$) is constituted by edge-sharing $CoO_6$ and $TeO_6$ octahedra that form a perfect honeycomb lattice of $Co^{2+}$ ions [Fig. 1(b)]. Such honeycomb layers are running perpendicular to the *c*- axis, and well separated from each other (~ 5.61 Å) by intermediate $Na^+$ layers. Although the magnetic honeycomb lattice is comprised by two crystallographically independent $Co^{2+}$ sites [Wyckoff positions 2*b* and 2*d*, respectively], the oxygen octahedra surrounding the sites are similar with the Co-O distances differing by only about 1% [15], resulting in almost identical environment. Anisotropic Kitaev interactions are predicted to exist between Co ions [5, 6] [Fig. 1(b)], in addition to Heisenberg interactions. Most interestingly, similar to $Na_2IrO_3$ and α-$RuCl_3$ [4, 25-27] a rich temperature-field phase diagram was also reported for $Na_2Co_2TeO_6$ [11, 25-30]. Such phase diagrams are regarded as a characteristic feature of the Heisenberg-Kitaev anisotropic model and reveal the presence of a possible QSL phase in the vicinity of the quantum critical point [31]. Although the magnetic properties of $Na_2Co_2TeO_6$ have been recently intensively studied [11, 14, 17, 18, 29, 30, 32-36], the question on the nature of field induced states (in particular, in regard to the proposed Kitaev QSL) remains open. Several recent studies [17, 18, 32, 36] reported the possible presence of the Kitaev QSL state in the intermediate fields above the field-induced transition at $H_c$, however, other studies [11, 33, 35] suggested the absence of the QSL state.

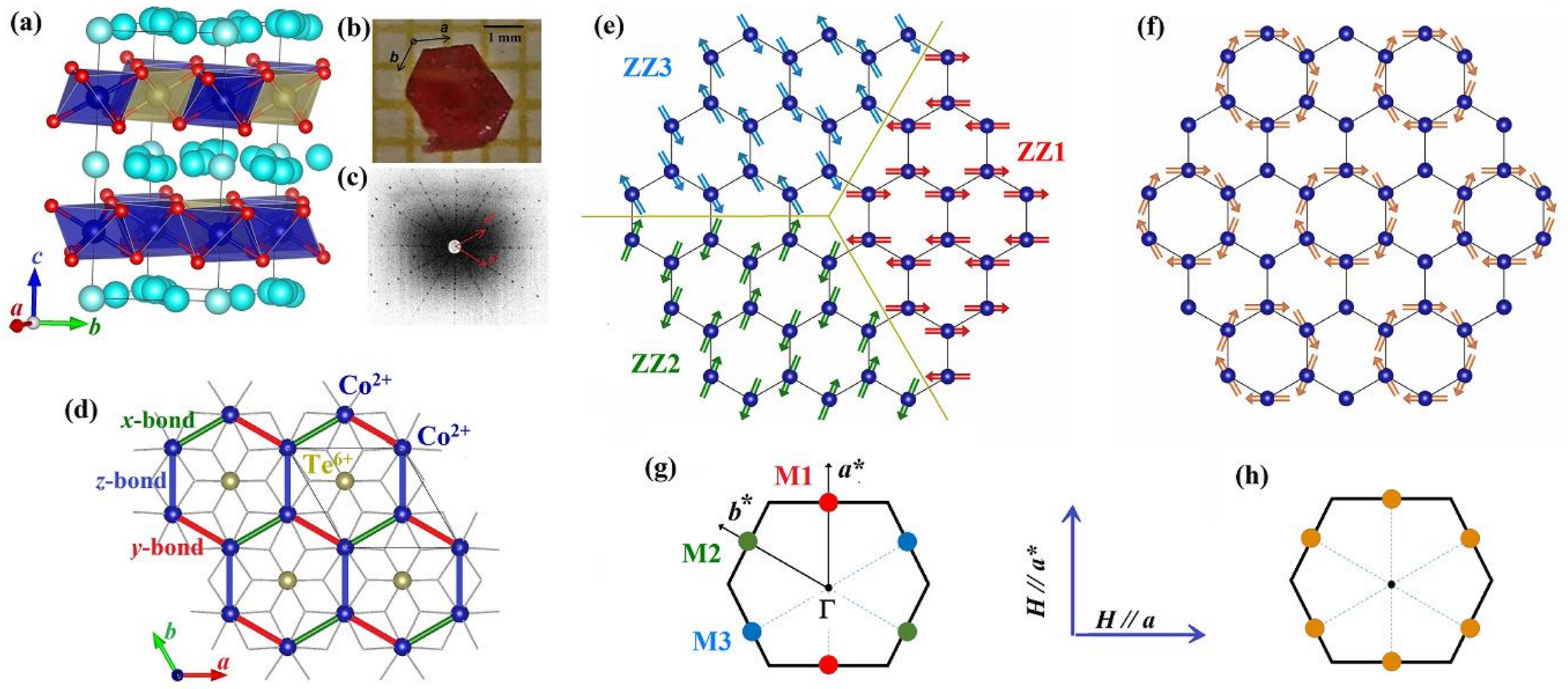

***FIG 1:*** *(a) The layered type crystal structure of $Na_2Co_2TeO_6$. (b) Photograph of a representative crystal on a millimeter grid. (c) The x-ray Laue diffraction pattern from the crystal. (d) The in-plane honeycomb layer of $Co^{2+}$ ions with the $Te^{6+}$ions in centers of honeycomb cells. Kitaev interactions between the x, y, and z components of the pseudospins are shown by thick lines with different colours. (e) A schematic of the collinear zigzag magnetic structure and its orientation domains. (f) A schematic of the noncollinear triple-Q magnetic structure. (g) and (h) The Bragg peak positions in the first Brillouin zone for the collinear zigzag AFM and non-collinear triple-Q structures, respectively. Magnetic structures were drawn by using VESTA 3 software [48].The scheme of the applied field directions in the present study is also shown.*

In this letter, we report results of an extensive neutron diffraction study of high-quality single crystals of $Na_2Co_2TeO_6$ for inplane applied magnetic fields with *H*//*a* and *H*//*a** directions (for the field direction schemes see Fig. 1)], which enable us to reveal the field evolution of its magnetic phases. Magnetic field dependent phase diagrams are proposed based on the neutron diffraction results. Furthermore, we report very different behaviors for the field evolutions of the magnetic phases for out-of-plane applied magnetic field (*H*//*c*) as compared to the inplane applied magnetic field for the Kitaev-Heisenberg material $Na_2Co_2TeO_6$.

*Experimental:* Polycrystalline samples of $Na_2Co_2TeO_6$ were prepared by solid-state method [15]. High-quality single crystals were grown by a self-flux method [14]. The crystals were characterized by x-ray backscattering Laue diffraction which confirms the six-fold symmetry [Fig. 1(b-c)] and the good quality of our crystals. The magnetization measurements on single crystal samples (for applied field along principal axes) were performed with a vibrating sample magnetometer (VSM, Cryogenic Co. Ltd., UK) [Fig. 1(b-c)]. Temperature dependent zero-field powder neutron diffraction measurements and magnetic field dependent single crystal diffraction measurements (for *H*//*a* and *H*//*a**) were performed on the WISH diffractometer at the ISIS facility, RAL, UK [37]. The measured zero-field powder diffraction patterns were analyzed using the Rietveld refinement technique, employing the FULLPROF computer program [38]. High-field (*H*//*c*) electron spin resonance (ESR) measurements were performed employing a 16 T transmission-type ESR spectrometer (similar to that described in Ref. [39]) in the 100 − 700 GHz frequency range; the experiments were done in the Faraday configuration.

*Results and discussion:*

The single crystals are preliminary characterized by temperature- and field-dependent magnetization measurements with magnetic field applied along three principal axes *H*//*a*, *H*//*a**, and *H*//*c* [Fig. S1], which reveal that the magnetism in $Na_2Co_2TeO_6$ is highly anisotropic and close to tipping points between competing phases. The observed results have been used to draw the phase diagrams [Fig. S1], which reveal field induced transitions at ~ 60 kOe for both *H*//*a* and *H*//*a**. The field-induced phases are proposed to host the Kitaev QSL state. The above results are in a good agreement with that reported earlier [11]. In the present study, we employed field-dependent single-crystal neutron diffraction to determine the field evolution of the magnetic phases at 1.5 K [the experimental geometry is shown in Fig. 1]. Before, presenting the results of our field-dependent single-crystal neutron diffraction study, we first describe below the magnetic ground state (in zero magnetic field) of the compound. The magnetic ground state of $Na_2Co_2TeO_6$ was initially reported to be a zigzag based collinear AFM structure by neutron powder diffraction [15, 16]. The long-range zigzag AFM ordering [with propagation vector $\boldsymbol{k}$= (1/2, 0, 0)] [15] manifests itself as sharp magnetic Bragg peaks at the *M*-points of the Brillouin zone (BZ) in the (H, K, 0) plane. As the zigzag AFM order breaks the threefold rotation symmetry (*C*3) of the hexagonal lattice, three magnetic domains (differing by an angle of 120°) are present [25]. Therefore, in the zigzag AFM phase [9, 15, 16], one should expect a total of six peaks [Fig. 1(g)] at six M-points (represented by the propagation vectors $\boldsymbol{k}_1$= (½ 0 0), $\boldsymbol{k}_2$= (0 ½ 0) and $\boldsymbol{k}_3$= (½ -½ 0), respectively) from the three magnetic domains. However, based on recent elastic and inelastic neutron scattering studies on single crystal samples as well as theoretical modeling on $Na_2Co_2TeO_6$ [9, 40, 41], it has been established that a "triple-*Q*" magnetic state [Fig. 1(f)] is most appropriate one. It is important to add here that the Bragg-peak pattern of a single magnetic domain of a triple-Q state (consists of six sharp peaks at all the M-points [Fig. 1(h)]), is identical to the averaged pattern of three domains of the single-*Q* zigzag AFM state. Therefore, the triple-*Q* and zigzag orders cannot be distinguished by neutron diffraction [9], unless the populations of the domains becomes unequal by external perturbations like magnetic field [40], strain, and/or pressure i.e., by breaking of the hexagonal *C*3 symmetry. Besides, a recent study on $Na_2Co_2TeO_6$ using muon spin relaxation [42] reveals the presence of prevalent spin dynamics with spatially uneven distribution and varied correlation times as well. The muon spin relaxation results imply that the magnetic ground state of $Na_2Co_2TeO_6$ cannot be solely described by the long-range triple-*Q* static order; suggesting a significant role of quantum fluctuations in establishing its ground state.

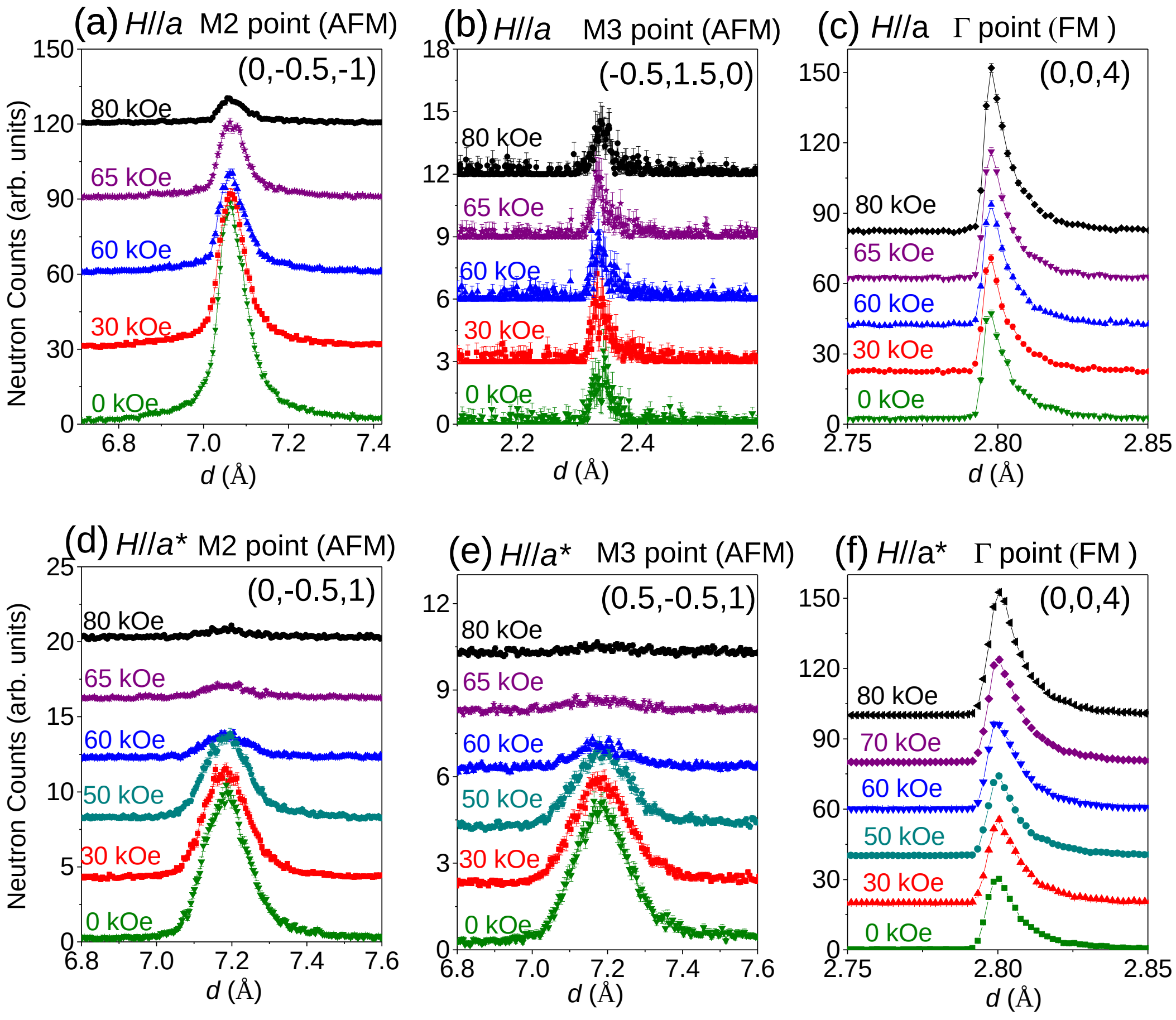

***FIG. 2:*** *The field evolution of representative magnetic Bragg peaks at the M2, M3, and Γ points of the BZ for (a-c) H//a and (d-f) H//a*, respectively. The measured diffraction patterns are offset vertically for clarity. The asymmetric peak shapes, especially for low-d value Bragg peaks, appear due to intrinsic feature of the time-of-flight diffractometer. Here, the moderation process generates pulses of neutron that, at a fixed wavelength, have a characteristic back-to-back exponential profile which in turn is responsible for the peak shape of the Bragg reflections.*

The field-induced evolution of the magnetic state is investigated directly by measuring the variations of magnetic Bragg peak intensities at the M points in the BZ as well as nuclear Bragg peaks at the Γ-point of the BZ as a function of applied magnetic field for both the applied field directions *H*//*a* and *H*//*a** [Fig. 2(a-f)]. The corresponding variations of the integrated intensities are shown in Fig 3(a-f), indicating anomalies at critical fields for both *H*//*a* and *H*//*a**. For the applied field along *H*//*a*, the intensities of all magnetic Bragg peaks [(0,-0.5,0), (0,-0.5,-1), (0,-0.5,-2), and (0,-0.5,-3)] corresponding to the M2 reciprocal points decrease linearly with the increasing field up to $H_{c1}$ = 60 kOe and attain ~ 40-60% of their corresponding zero-field intensities [Fig. 3(a)]. Above $H_{c1}$, a linear decrease of the intensities is also evident. The slope ($dI/dH$) is higher for $H > H_{c1}$ than that for the $H < H_{c1}$ [Fig. 3(a)]. The change of slope occurs at $H_{c1}$ in line with the bulk magnetization [Fig. S1]. A similar behavior has also been found for the magnetic Bragg peak at the M3 point [Fig. 3(b)] of the second BZ. It may be noted here that the variation of the intensity for the magnetic Bragg peaks with *L*=odd indices is slightly larger than that for the *L* = even case. Due to the instrumental geometry with the vertical magnetic field, no magnetic Bragg peaks corresponding to the M1 point of the BZ are accessible. In addition, we have shown the intensity variation of the selective nuclear Bragg peaks at the Γ-point [Fig. 3(c)]. An enhancement of the intensities of the nuclear Bragg peaks (0,0,4) and (0,1,0) is found with increasing magnetic field. The additional magnetic signal at these nuclear Bragg peak points is, therefore, due to a field induced uniform magnetization. This means that the suppression of the antiferromagnetic order is due to the significant spin polarization along the applied magnetic field *H*//*a*. However, the maximum applied field of 80 kOe is not sufficient to achieve the complete parallel alignment of the moments as also indicated by the bulk magnetization data. This implies that the system is in a magnetically ordered state under applied field *H*//*a* up to 80 kOe, and the expected field-induced spin liquid state is not yet achieved (discussed in details later).

Now, we present additional data for the in-plane applied field direction *H*//*a**. The variation of the intensity of the magnetic peaks with the applied magnetic field reveals a completely different behavior than that for the *H*//*a*. With the increasing field for *H*//*a**, the intensities of the magnetic Bragg peaks at M2 and M3 points enhance slightly for H up to ~ 5 kOe, and then decrease linearly over 5 < *H* < 57 kOe (*H*c1) [Fig. 3(d-e)]. However, at *H* = *H*c1, the intensities of the magnetic Bragg peaks with *L* = odd indices [i.e., (0,-0.5,1), (0,-0.5,3), (0.5,-0.5,1) and (0.5,-0.5,5)] fall sharply by ~40 %, and attain ~ 20% of their zero field values. The observed sharp fall is in contrast to the results for *H*//*a*, presented in the previous section. Nevertheless, they are in-line with the observed first-order field-induced transition of magnetization [Fig. S1]. In contrast, no sharp drop in intensity at the *H*c1 is observed for the magnetic Bragg peak with *L* = even indices [i.e., (0,-0.5,4) and (-0.5,-1.5,0)], rather only a change in slope ($dI/dH$) is evident at *H*c1 (as like the case for *H*//a). Upon further increasing the magnetic field above the *H*c1, the intensities decrease linearly. The residual intensity of the magnetic Bragg peak (0,-0.5,4)] with *L* = even is found to be ~ 50 % of the zero field intensity at 80 kOe as compared to ~ 10 % for the magnetic peaks with *L* = odd indices [i.e., (0,-0.5,1) and (0,-0.5,3)]. Such contrasting behaviors for the planes with *L* = odd and *L* = even are unique and have been revealed for the first time by the present comprehensive single crystal neutron diffraction study. This is consistent with the antiferromagnetically coupling of the magnetic planes having triple-*Q* structure along the *c* axis in $Na_2Co_2TeO_6$ [15] and reveals a different chirality for the alternating magnetic planes along the *c* axis. For this field configuration as well, an enhancement of the intensities of the nuclear Bragg peaks (0,0,4) and (-1,2,0) are found with increasing magnetic field [Fig. 3(f)]. However, the relative change of the intensity is much stronger as compared to the same for *H*//*a*. As the neutron diffraction is only sensitive to the ordered moment component perpendicular to the wave-vector transfer $\boldsymbol{Q}$, the enhancement of the intensity of the (0,0,4) nuclear Bragg peak suggests that the FM component lies mainly within the ab plane. The in-plane FM component is also evident from the higher bulk magnetization values for the in-plane fields (*H*//*a* and *H*//*a**) as compared to that for the *H*//*c* (Fig. S1). Further, the enhancement of the intensities of the Bragg peaks (-1,2,0) for *H*//*a** reveals that the FM moments develop along the applied field directions. Therefore, the additional magnetic signal of the nuclear Bragg peaks is due to field induced uniform magnetization as like the case for *H*//*a*. The intensities of the AFM Bragg peaks persist (~ 20-40% of their corresponding intensities in the zero-field) up to 80 kOe for *H*//*a** revealing the presence of the magnetically ordered state. This also implies that the spin liquid state is not achieved even up to a field of 80 kOe for *H*//*a** as well (discussed in details later). Nevertheless, our results unambiguously reveal different field dependencies for the *H*//*a* and *H*//*a** directions especially around the critical magnetic field *H*c1 as evident from the direct comparison of the field dependence for the M2 (0,-0.5,1) point [Fig. 3(g)]. Such a different field dependence is unusual considering the *C*3 symmetry of the recently proposed triple-*Q* magnetic ground state.

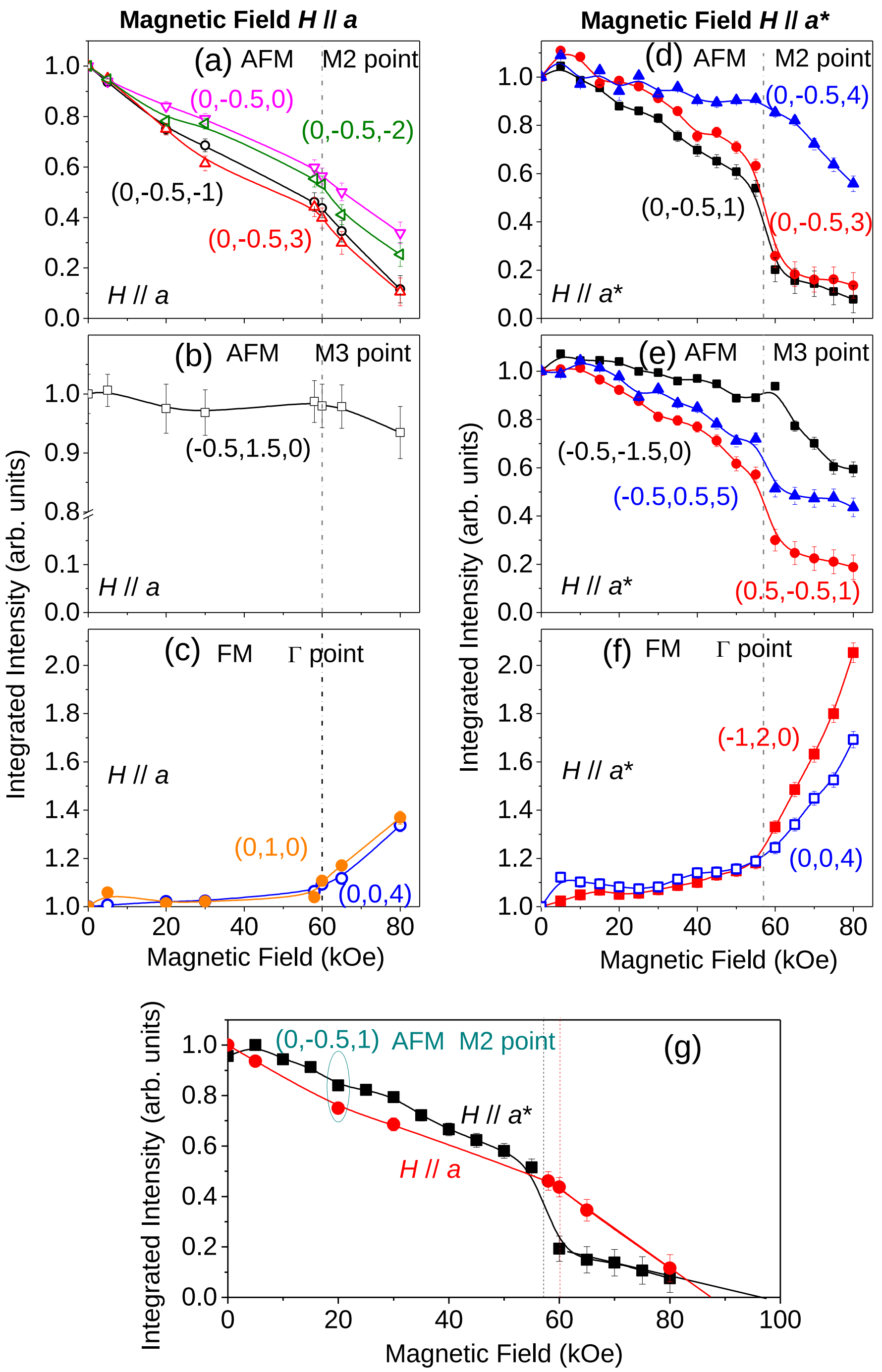

***FIG. 3:*** *(a-c) The field evolution of the integrated intensity of magnetic Bragg peaks of* $Na_2Co_2TeO_6$ *at M2, M3 and Γ points of the BZ under the magnetic fields for (a-c) H//a and (d-f) H//a*, respectively. . All the intensities of the magnetic Bragg peaks are normalized with respective to their zero field values. (g) A comparison between the field dependence of the intensity of the AFM peak at M2 point for H//a and H//a*.*

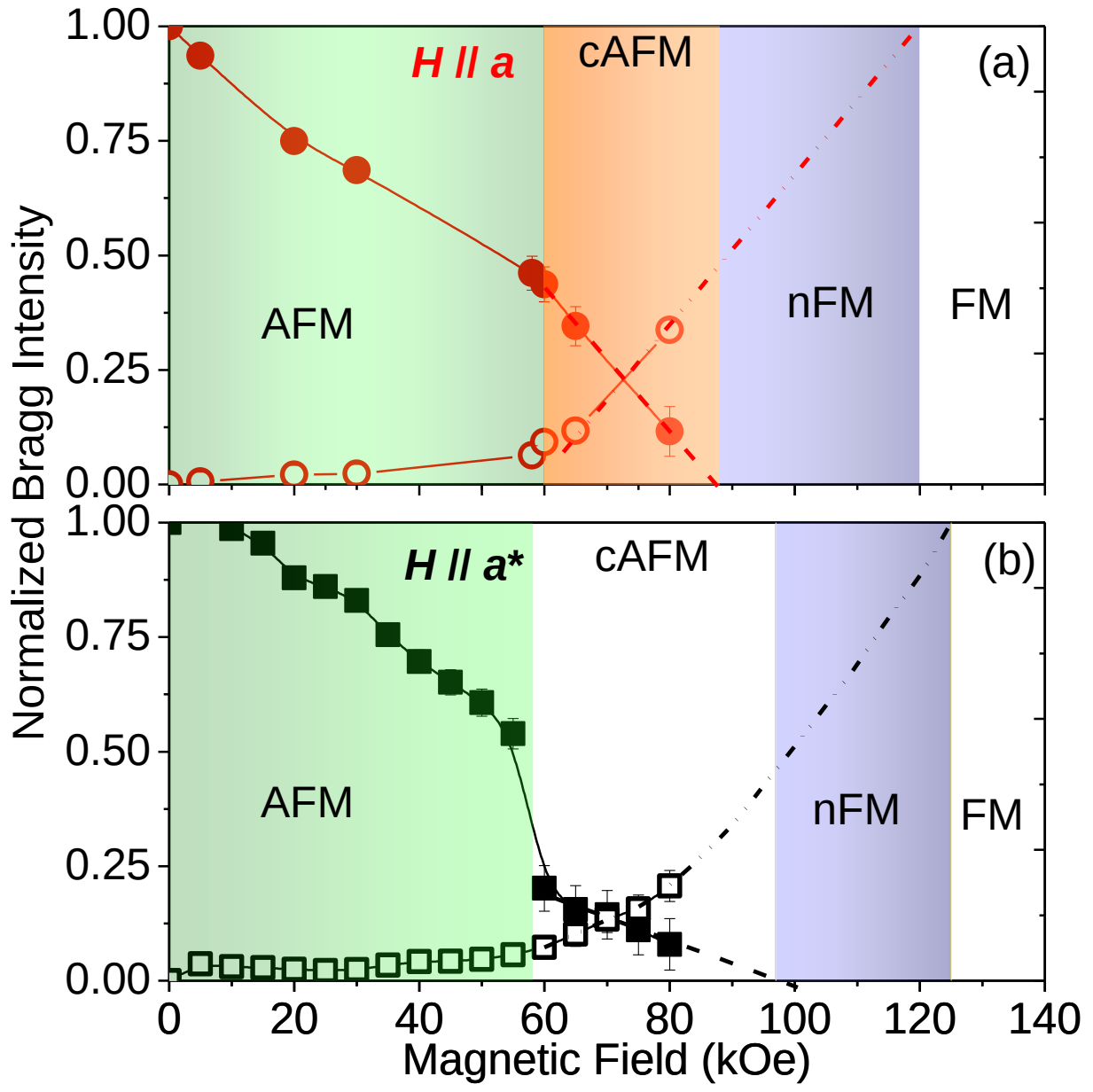

***FIG. 4***: *The magnetic phase diagrams of* $Na_2Co_2TeO_6$ *at 1.5 K for (a) H//a and (b) H//a*, respectively. The AFM, cAFM, nFM, and FM stand for antiferromagnetic, canted antiferromagnetic, near field polarization, and field polarized states, respectively.*

The observed variations of the normalized intensity of the AFM and FM components from the present single crystal neutron diffraction study at 1.5 K are fruitfully used to draw the field-dependent magnetic phase diagrams for the *H*//*a* and *H*//*a** [Fig. 4]. For the AFM Bragg phase, the peaks intensity is normalized to the zero-field intensity. The intensity of the nuclear Bragg peak (0,0,4) is normalized with the intensity expected for the completely polarized (FM) state with the magnetic moment along the *a* and *a** axes corresponding to the cases for *H//a* [Fig. 3(c)] and *H//a** [Fig. 3(f)], respectively. The linear extrapolation of the AFM intensity reveals that it persists up to $H_{c2}$ ~ 90 and 95 kOe for the applied fields *H*//*a* and *H*//*a**, respectively [Fig. 4]. On the other hand, the extrapolation of the FM intensity reveals that the full polarization occurs only above $H_S$~ 120 and 125 kOe for the applied fields *H*//*a* and *H*//*a**, respectively. This agrees with the bulk magnetization where a near saturation is achieved above ~ 120 kOe. In agreement with our results, recent magnetic phase diagrams have been proposed for $Na_2Co_2TeO_6$ under in-plane applied fields, suggesting the existence of two field-induced intermediate states (prior to reaching the field-polarized state) [43]. The magnetic phases between the $H_{c1}$ and $H_{c2}$, where the coexistence of both the AFM and FM Bragg peaks are observed, can be labelled as canted antiferromagnet (cAFM). Recent theoretical study predicts a metamagnetic phase transition (in the classical limit) from the triple-Q ground state to a canted zigzag state (z-zigzag for *H*//*a** and *x*/*y*-zigzag for *H*//*a*) at intermediate fields, preceding the transition toward the field-induced polarized state at higher fields [41]. A comparison of the measured intensities (Fig. 3) for both the field directions with the simulated magnetic Bragg peak intensities (considering such cAFM structures with the moment components for 80 kOe) gives a satisfactory agreement [Fig. S5 (a-b)]. Further analysis of our single crystal data [Fig. S6] rules out the presence of any additional field induced ordered state (as reported for the most extensively studied Kitaev-Heisenberg model compound *α*-$RuCl_3$ and labelled as 2nd zig-zag AFM phase, ZZ2 state) [44]) over this intermediate field range. Moreover, the phase diagrams show that only a partial polarized (FM) phase is present over the field region $H_{c2}$-$H_S$.

Our recent high-resolution terahertz spectroscopy study on $Na_2Co_2TeO_6$ as a function of applied magnetic field with different terahertz polarizations [45] illustrate spin dynamics with distinct characteristics over magnetic field 0-70 kOe, 70-100 kOe and above 100 kOe. While below 70 kOe and above 100 kOe the dynamics is characterized by well-defined magnetic excitations, in the intermediate regime 70-100 kOe reveals sharp absorption profile and extended continuum in the longitudinal and transverse polarization channels for both the applied field directions *H*//*a* and *H*//*a**. A polarization-selective continuum in the intermediate phase are of indication for spin fluctuations of a proximate quantum spin liquid. Furthermore, recent experimental studies of $Na_2Co_2TeO_6$ employing various techniques, such as, magnetic torque [29], specific heat [17], thermal transport [18, 33], have unveiled intriguing properties under in-plane fields above the critical field. These observations imply that the field-induced near-polarized states exhibit characteristics distinct from those of a conventional paramagnet. Additionally, a recent theoretical study has reported a field-induced transition from the low-field ordered state to a gapless quantum spin liquid at intermediate fields, occurring prior to the emergence of the field-induced polarized state [46].

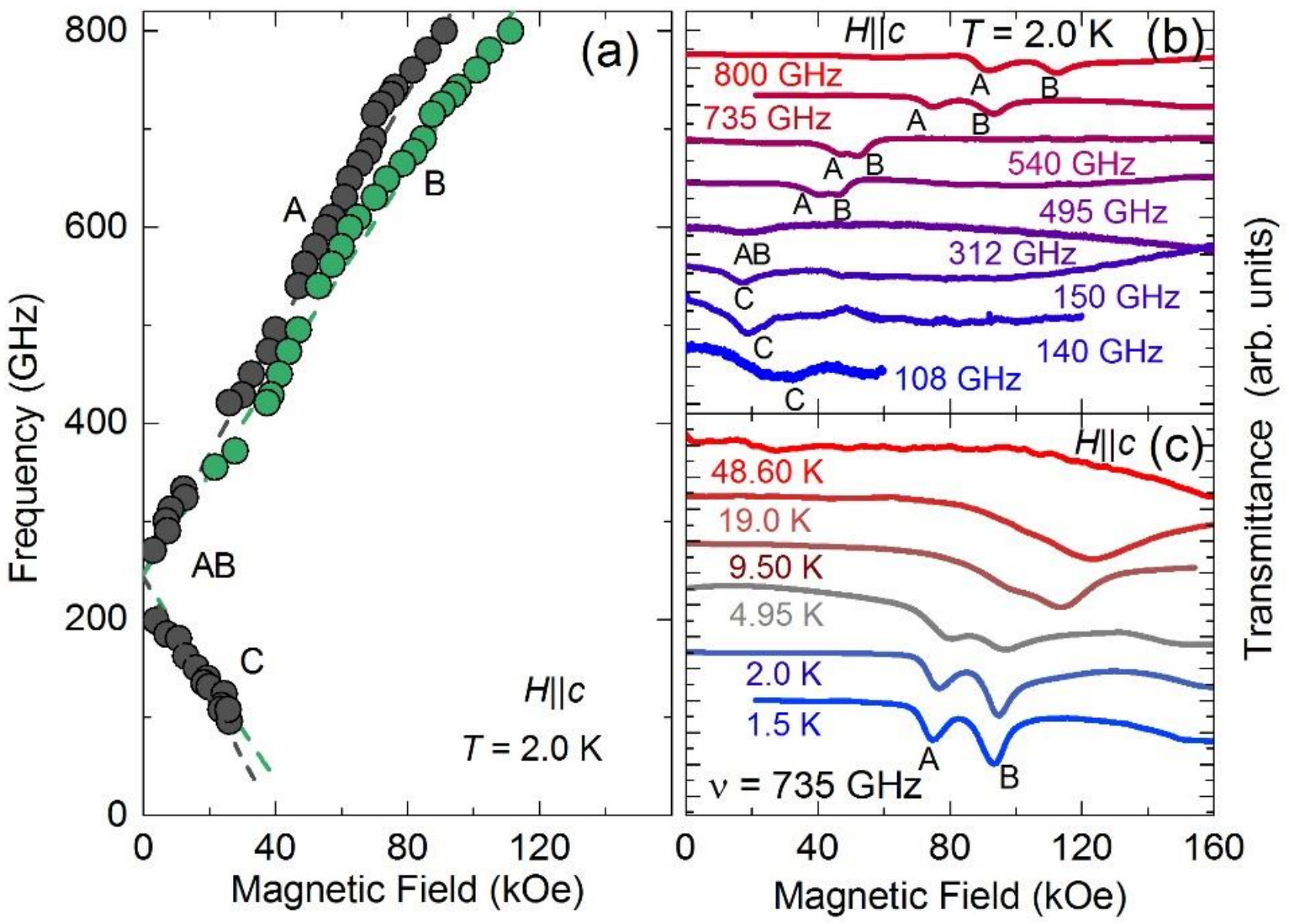

***FIG. 5:*** *(a) Frequency-field dependences of ESR excitations measured at T = 2.0 K with magnetic field applied along the c axis. The dashed lines are guides to the eyes. (b) Examples of ESR spectra taken at T = 2.0 K. (c) ESR spectra measured at a frequency of 735 GHz and different temperatures. The spectra in (b) and (c) are offset vertically for clarity.*

Now we shed light on the nature of the field dependent state for *H//c* by field and temperature dependent electron spin resonance (ESR) measurements on the $Na_2Co_2TeO_6$ single crystals. The ESR spectra measured at 2 K for *H*//*c* [Fig. 5(a)] reveal three modes of antiferromagnetic resonance (AFMR) *A*, *B*, and *C*, respectively. The extrapolation of the frequency-field dependences of the observed excitations to zero field suggests the presence of the energy gap, $\Delta$ = 240 GHz (~ 1 meV), which is similar to that observed previously by inelastic neutron scattering [30, 47]. The modes *A* and *B* demonstrate linear increase in the resonance-frequency position, while mode $C$ exhibits softening with increasing field. The extrapolation of the $C$ mode suggests a complete softening at ~ 40 kOe where a change in slope in the *MvsH* curve as well as a change in the nature of the susceptibility curves is found (Fig. S1). The above experimental observation might be an indication of a new field induced phase transition at 40 kOe. Upon increasing the temperature, the AFMR modes $A$ and $B$ become broader and cannot be resolved anymore at or above $T_N$ = 27 K [Fig. 5]. The mode $C$ exhibits a substantial broadening at low frequencies, suggesting the presence of two resonance lines (AFMR modes $C$ and $C'$), similar to that revealed for the higher-frequency absorptions (AFMR modes $A$ and $B$). The presence of this double-peak structure of the ESR absorptions can be explained taking considering possible tiny twinning of the crystal along the *c*-axis.

*Conclusion:* In conclusion, we have presented results of single crystal static magnetic susceptibility, neutron diffraction, and ESR studies along with powder heat capacity of $Na_2Co_2TeO_6$, allowing us to shed light on the microscopic nature of the field–induced phases. Our comprehensive study under applied magnetic field along *H*//*a*, *H*//*a** and *H*//*c* determines the field-evolution of the magnetic phases. Our findings reveal that phase transitions occur from the ordered ground state to an ordered canted zigzag antiferromagnetic state at approximately 60 kOe, followed by a possible transition to a partially polarized state (over the range of approximately 90-120 kOe), and ultimately to a field-induced fully polarized state (above approximately 120 kOe). Notably, our results also highlight the distinct field dependencies of the magnetic peak intensities for in-plane field directions *H*//*a* and *H*//*a**. These field-induced near-polarization states exhibit distinct characteristics and warrant further investigation to unveil their exotic magnetism. Conversely, for *H*//*c*, we have identified a new phase transition at 40 kOe through field and temperature-dependent electron spin resonance (ESR) measurements.

**Acknowledgment:**

A.K.B. and S.M.Y. thank the Department of Science and Technology, India (SR/NM/Z-07/2015) for the access to the experimental facility and financial support to carry out the neutron scattering experiment, and Jawaharlal Nehru Centre for Advanced Scientific Research (JNCASR) for managing the project. The authors thank the Science and Technology Facilities Council (UK) for the provision of neutron beam-time on the WISH beam line (https://doi.org/10.5286/ISIS.E.RB2010435-1 and https://doi.org/10.5286/ISIS.E.101136877). For the purpose of open access, the author has applied a creative commons attribution (CC BY) license to any author accepted manuscript version arising. Data are available under reasonable request from the authors. This work was supported by the Deutsche Forschungsgemeinschaft and SFB 1143, as well as by HLD at HZDR, member of the European Magnetic Field Laboratory (EMFL).

# Supplementary Information:

### (i) Bulk magnetization and *H*-*T* phase diagram:

The temperature dependent ZFC and FC susceptibility (*M/H*) curves for *H*//*a*, *H*//*a**, and *H*//*c* measured under a magnetic field of 0.5 kOe using the single crystal are shown in Fig. S1(a). The magnetic ordering at $T_N$ is manifested as a peak like anomalies in all the curves. With further decreasing temperature, additional anomalies are found at ~ 20 and 7 K in agreement with the several reports [1, 2]. The magnetic field dependences of the ZFC susceptibility for all the three field directions *H*//*a*, *H*//*a**, and *H*//*c* are shown in Figs. S1(b, c, and d), respectively. With increasing field, $T_N$ decreases continuously for *H*//*a*, and *H*//*a**, whereas, it remains almost unchanged for *H*//*c*. Moreover, for *H*//*a*, and *H*//*a**, the low temperature anomalies get strongly suppressed with increasing magnetic field up to 10 kOe, and disappear completely above ~ 60 kOe. Whereas for *H*//*c*, the low temperature anomalies are nearly suppressed above 1 kOe. Moreover, a change in the nature of the susceptibility curves is evident above 40 kOe [inset of Fig. S1(d)], revealing a field induced phase transition as confirmed by our ESR data [Fig. 5].

The isothermal magnetization (M vs. *H*) curves measured at 2 K reveal a field induced transition at $H_c$ ~ 57 and 60 kOe for *H*//$a^*$ and *H*//*a*, respectively [Fig. S1(e)]. For *H*//$a^*$, a hysteresis is found at around the critical field $H_c$ ~ 57 kOe between the up and down sweeps of the magnetic field in the M vs. *H* study, suggesting a first order field-induced transition at $H_c$. For *H*//$a^*$, with the increasing temperature the $H_c$ value decreases slightly [Fig. S1(f)]. Moreover, the hysteresis around $H_c$ in the *M*(*H*) curve gradually decreases and disappears above ~ 20 K. On the other hand, for *H*//*c*, the *M*(*H*) curve shows a much smaller magnetization value as compared to the *H*//$a^*$ and *H*//*a* cases. Interestingly, *M*(*H*) for *H*//*c* shows near linear behaviour with a change of slope at ~40 kOe which corresponds to a field induced phase transition at ~40 kOe as also revealed by our ESR study. A saturation magnetization ~ 2 $\mu_B/Co^{2+}$ was confirmed in our *M*(*H*) study measured up to 140 kOe on a powder sample which is in good agreement with the reported value [3]. The above results thus reveal that the magnetism in $Na_2Co_2TeO_6$ is highly anisotropic and close to tipping points between competing magnetic phases. The observed results have been fruitfully used to draw the phase diagrams, depicted in Figs. S1(h) and S1(i), for *H*//*a** and *H*//*c*, respectively. The horizontal boundaries in the phase-diagrams are obtained from the anomalies in the *M*/*H* vs *T* curves [ Fig. S1(a)]. The vertical phase boundaries are obtained from the peaks in the *dM/dH* vs *H* curves [Fig. S1(g)]. The above results are in good agreement with that reported earlier [4] and confirm the good quality of our crystals.

### (ii) Field dependent Specific heat:

The temperature dependent heat capacity curves measured under different applied magnetic fields are shown in Fig. S2(a). The magnetic ordering at $T_N$=27 K (as found in the ZFC and FC magnetic susceptibility curves) is manifested as a sharp λ-like anomaly in the zero-field magnetic specific heat curve measured on the polycrystalline sample. With increasing field, the sharp peak at $T_N$ in the magnetic specific heat curves [Fig. S2(b)] (obtained after subtraction of the lattice contribution which was estimated by Debye–Einstein (DE) model) shifts toward lower temperatures and becomes broader. At 90 kOe, the peak becomes indiscernible which was referred to a magnetically disordered state due to the suppression of the magnetic ordering by Lin *et al*. [3]. Further, we have found about 27% of the theoretical total magnetic entropy remains as residue at 2 K in agreement with Ref. [3]. With the increasing field, especially above the $H_c$ (~ 57 - 60 kOe), in contrast to the conventional magnet, the residual entropy of $Na_2Co_2TeO_6$ increases and becomes about 36 % for

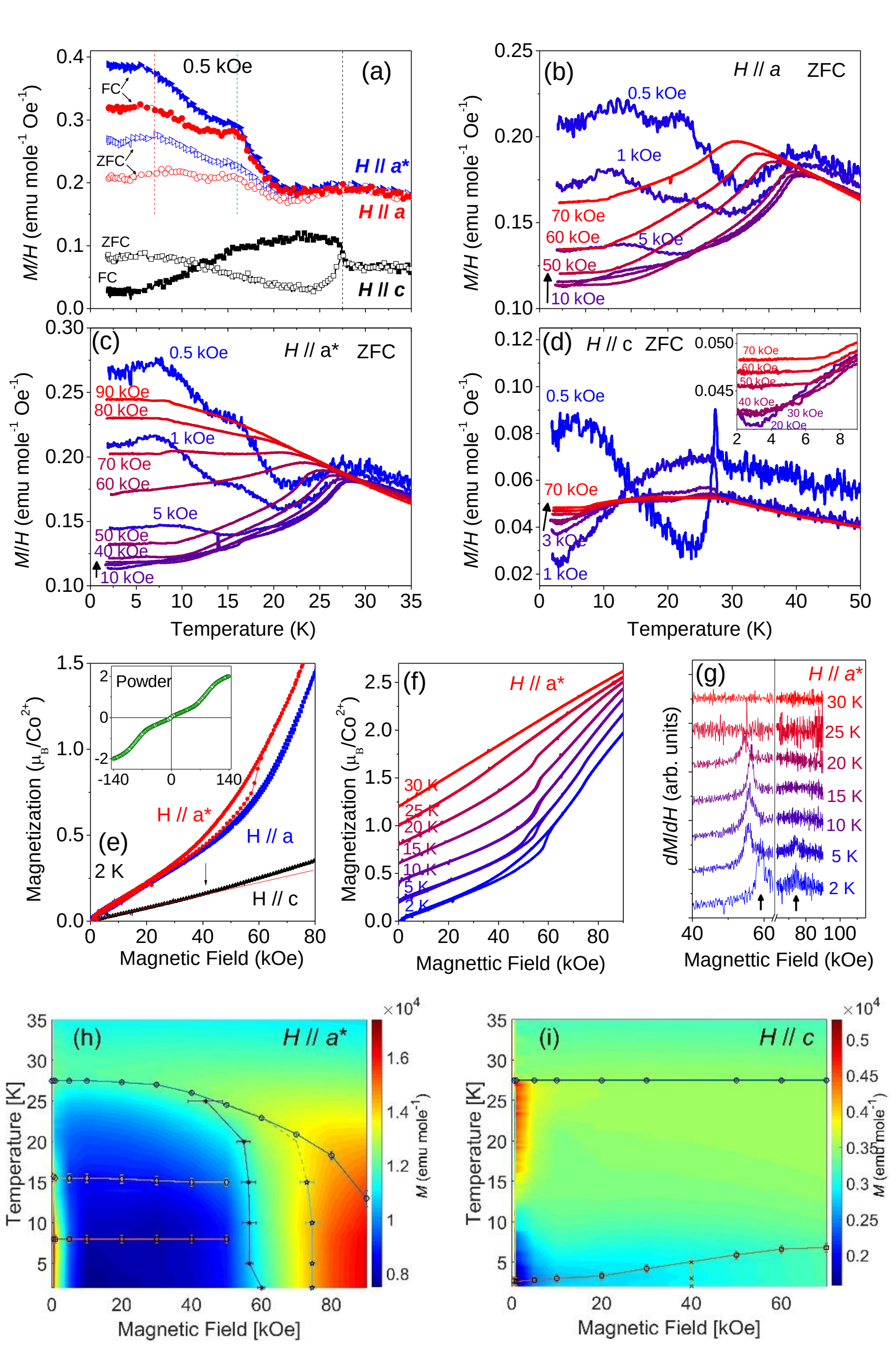

***FIG. S1:*** *(a) The temperature dependent ZFC and FC susceptibility (magnetization M divided by H) curves under 0.5 kOe of magnetic field along the crystallographic a*, a and c directions. The dashed lines represent the anomalies and used in (g) for the phase boundaries. (b)-(d) The magnetic field dependent ZFC susceptibility curves along the crystallographic a, a*, and c directions, respectively. The inset of (d) shows an enlarged view of low temperature region for field range 20-70 kOe, revealing the change of the nature of the M/H(T) curves above 40 kOe. (e) The isothermal magnetization curves M(H) as a function of magnetic field at 2 K along the a, a* and c directions. The inset shows the M(H) curves of powder sample up to 140 kOe. (f) The M(H) curves at different temperatures for H//a*. (g) The first derivatives of the dc magnetization curves [shown in (f)] with respect to magnetic field (dM/dH) as a function of magnetic field for H//a*. Two anomalies in the dM/dH vs H curves at ~ 60 kOe and 74 kOe are marked with arrows. (h-i) The magnetic phase diagram in the H-T plane for H//a* and H//c, respectively.*

130 kOe. These results strongly reveal the appearance of the field induced spin-disordered state. Several recent articles [3] reported that the zero-field AFM state coexists with a field induced spin disordered state above $H_c$.

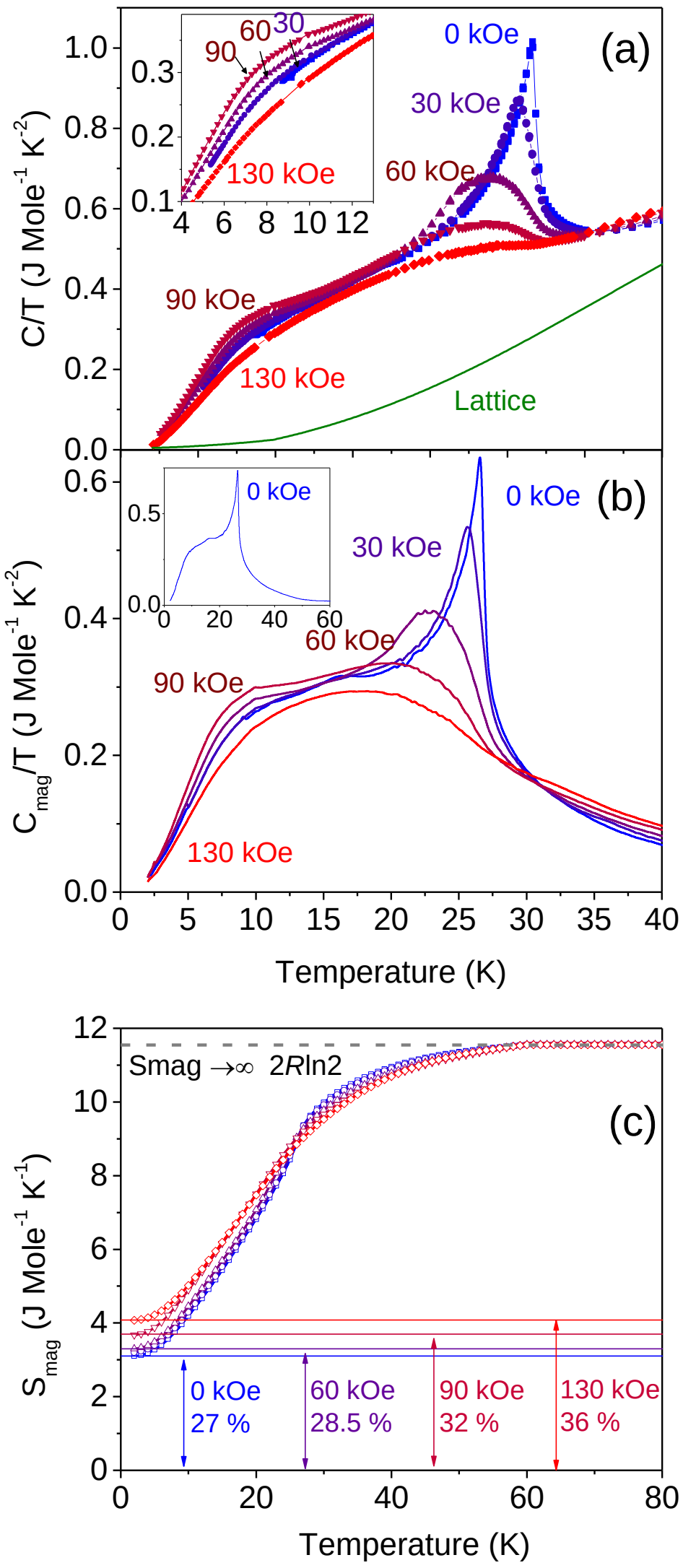

***FIG. S2:*** *(a) The temperature dependence of the specific heat of $Na_2Co_2TeO_6$ for magnetic fields 0, 30, 60, 90, and 130 kOe. The solid green line shows the lattice contribution of the specific heat, as calculated by Einstein-Debye model. The inset shows an enlarge view over the low temperature range. (b) The magnetic specific heat (Cm) as a function of temperature after subtraction of the lattice contribution. Inset shows the Cm vs T curves in zero field. (c) The temperature dependent magnetic entropy Smag (T) curves. The grey dashed line refers to Smag(T→∞) calculated with effective spin $J_{eff}=1/2$ for $Co^{2+}$. The Smag(T) curves are shifted vertically to coincide their maxima with the Smag(T→∞). Horizontal lines, respectively show the residual magnetic entropy at 2 K for 0, 60, 90, and 130 kOe.*

## (iii) Temperature dependent zero-field neutron powder diffraction:

Figure S3 shows the experimental magnetic neutron powder diffraction pattern, measured on WISH diffractometer, at several temperatures between 1.5 and 30 K. The temperature variations of the intensities of the most intense magnetic Bragg peaks (0,0,0)+***k*** and (0,0,1)+***k*** located at *d*= 9.2 and 7.1 Å, respectively, reveal the additional anomalies at $T_1$ = 16 and $T_2$ = 5 K (in agreement with the low field bulk susceptibility curves shown in Fig. S1) in addition to the one at $T_N$ = 27 K [Fig. S4(d)]. The temperature dependent lattice constants and unit cell volume show strong anomaly at $T_N$, however, no observable anomalies at the $T_1$ and $T_2$ were detected [Fig. S4(a-c)]. This suggests that the anomalies in the magnetization curves at $T_1$ and $T_2$ are possibly associated with small reorientations of ordered moments as outlined in Ref. [5].

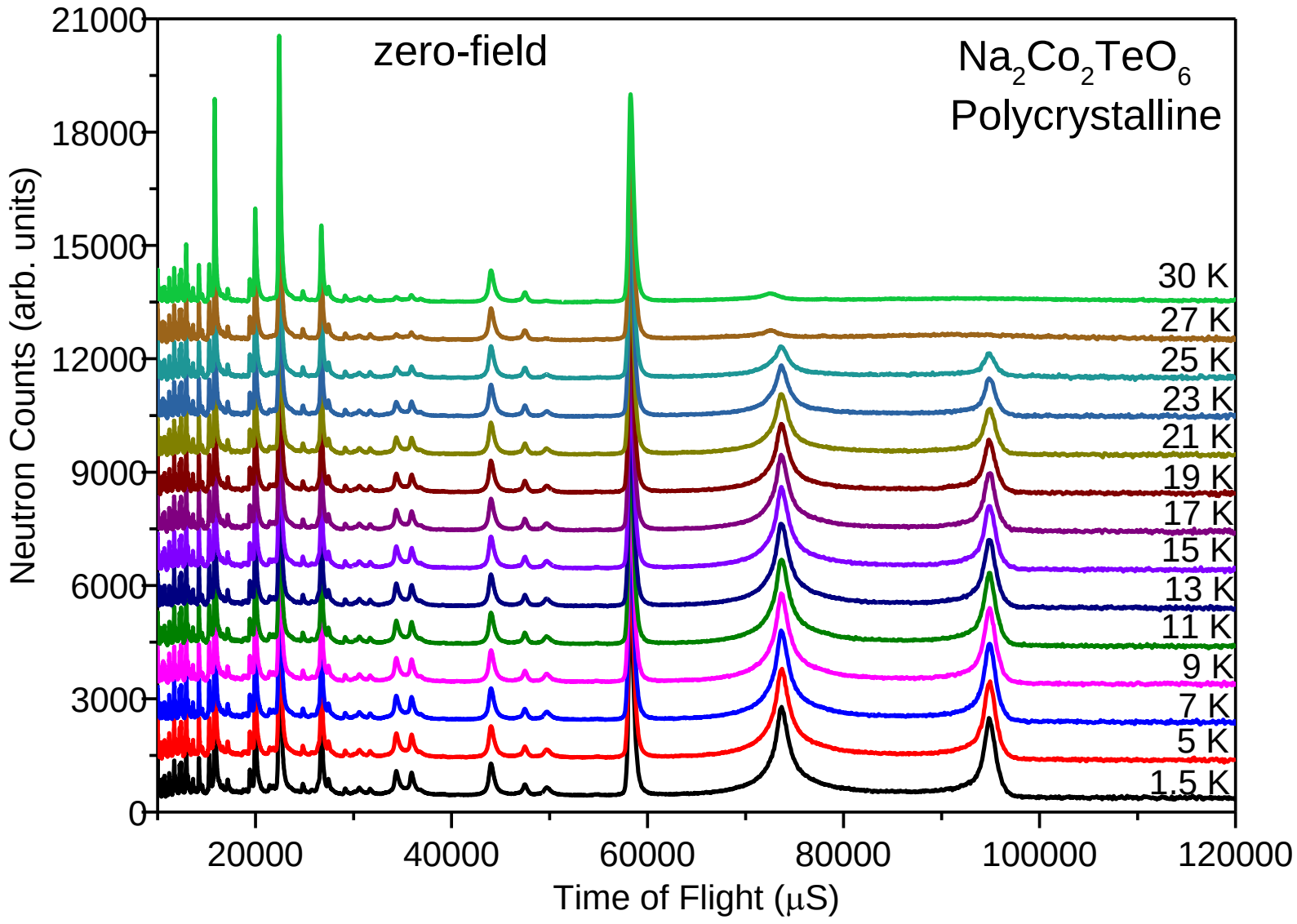

***FIG. S3:*** *(a) The experimentally measured powder neutron diffraction pattern of $Na_2Co_2TeO_6$ at several temperatures between 1.5 and 30 K in zero field.*

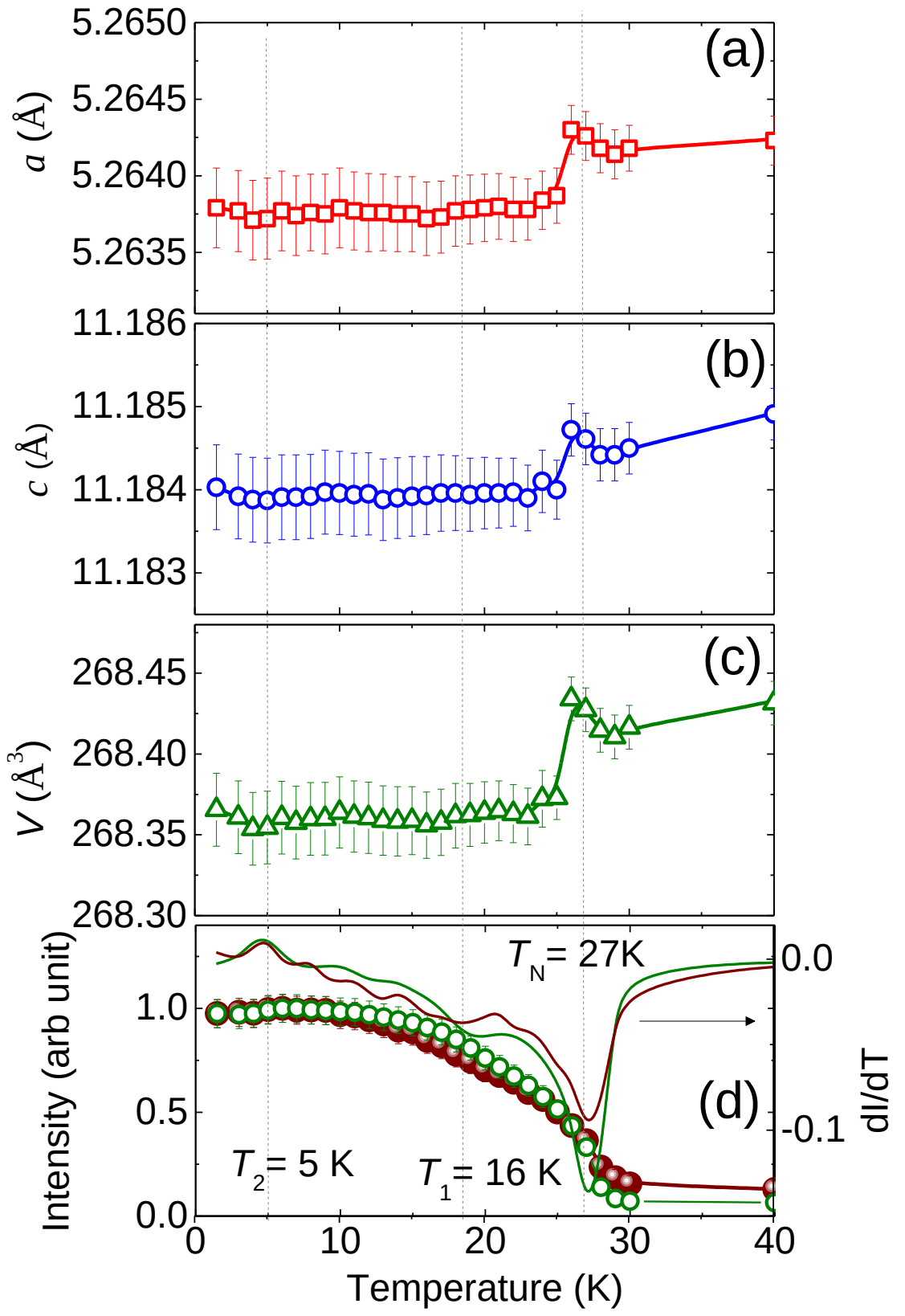

***FIG. S4:*** *The temperature dependence of the (a-b) lattice constants, (c) unit cell volume and (d) the integrated intensity of the strongest magnetic Bragg peaks (½,0,0) and (½,0,1) that are located at d =9.2 and 7.1 Å, respectively, in the powder neutron diffraction pattern of $Na_2Co_2TeO_6$ [Fig. S3].*

### (iv) Comparison of experimental and calculated intensities in the canted AFM state:

For the intermediate field range, the observation of the magnetic intensity at the fundamental nuclear Bragg positions (*k*=0) together with the decrease of the AFM reflections is a good indication of a canted AFM (cAFM) state. To further shed light on the cAFM state, we have calculated integrated intensity of the selected magnetic Bragg peaks (that are found experimentally) for a canted zigzag magnetic state model and compared the intensities with the experimentally measured intensities at 80 kOe (Fig. S5) for both *H*//*a* and *H*//*a**. For the calculation, we have considered the FM and AFM moment components of 1.93 and 1.24 for *H*//*a* and 1.63 and 1.24 for *H*//*a** considering the net ordered moment of 2.77 $\mu_B$/$Co^{2+}$, as determined in zero field. Good agreements of the calculated and experimental intensities are evident for both the field direction confirming a cAFM state.

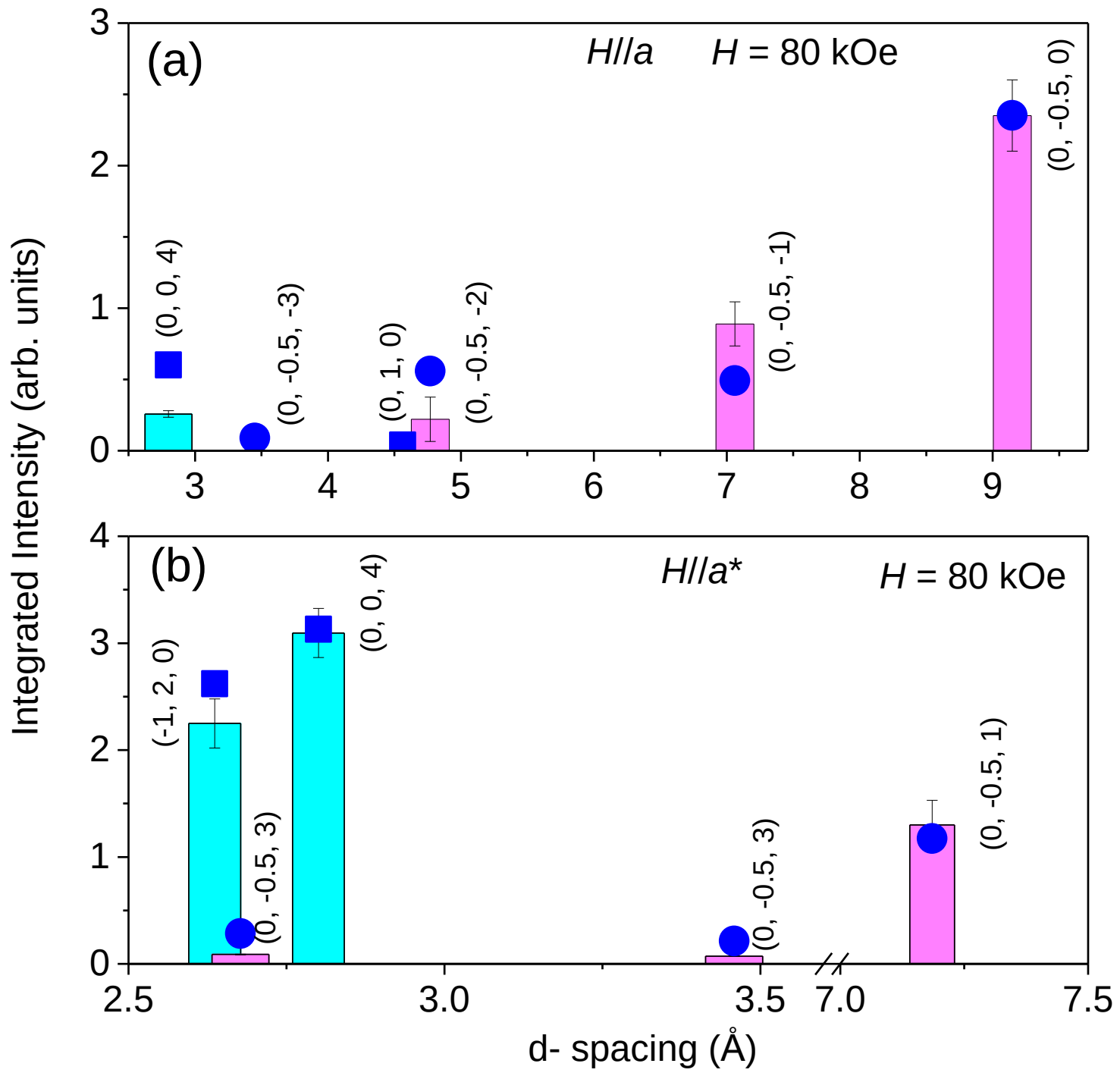

***FIG. S5:*** *A comparison of the experimental (measured under a magnetic field of 80 kOe) and calculated (based on the cAFM structures) intensities of magnetic Bragg peaks for (a) H//a and (b) H//a*. Height of the magenta and cyan colored bars, respectively, represent the measured magnetic Bragg peak intensities at the pure magnetic Bragg peak and fundamental nuclear Bragg peak positions. The Bragg peak intensities have been normalized by incident flux and Lorenz correction. The calculated magnetic intensities are shown by the filled circles and squares for the pure magnetic Bragg peaks and fundamental nuclear Bragg peaks, respectively.*

### (v) Field dependent neutron diffraction patterns at *M* and *Γ*-point:

The field-induced second zig-zag AFM phase or ZZ2 state for $\alpha$-$RuCl_3$ was characterized by a double periodicity in the direction perpendicular to the honeycomb planes where additional magnetic Bragg peaks appear at half integer values of *l*. In contrast, the diffraction patterns of the studied compound $Na_2Co_2TeO_6$ at 80 kOe do not show any additional magnetic peak revealing the absence of such field induced ordered state.

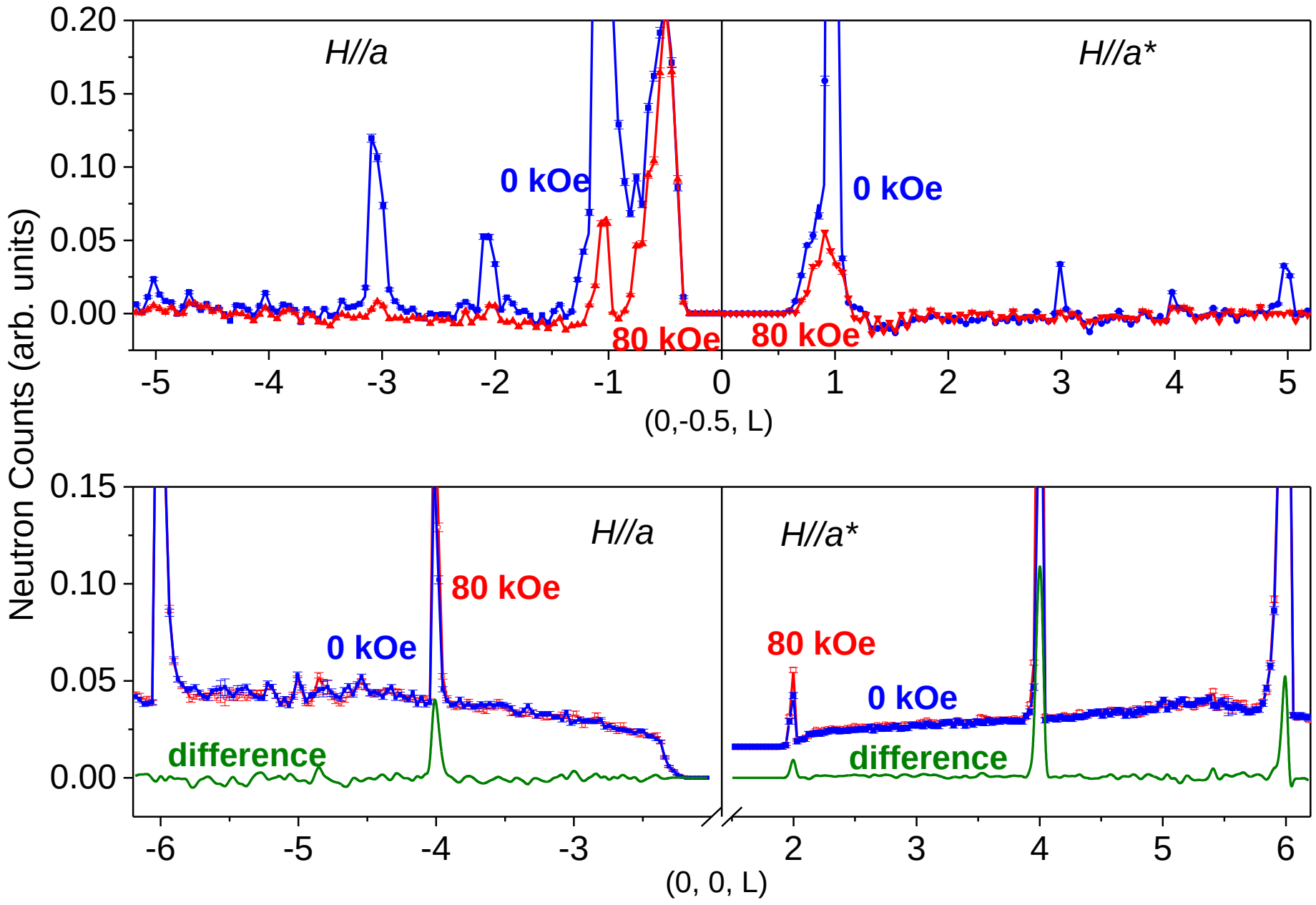

***FIG. S6:*** *(a) M-point (0,-1/2, L) and (b) Γ-point neutron-diffraction intensities as a function of the out-of-plane wave-vector component L for H = 0 kOe (blue) and 80 kOe (red), respectively. The data are integrated in ΔH =±0.05 r.l.u. and ΔK =±0.05 r.l.u. No additional reflection has been found in the field induced state at H~ 80 kOe.*